%% file: main.tex
\documentclass[aps,prl,reprint,groupedaddress]{revtex4-2}

\usepackage{graphicx}
\usepackage{dcolumn}
\usepackage{bm}

\usepackage[utf8]{inputenc}
\usepackage[T1]{fontenc}
\usepackage{booktabs, array, mathptmx, float, tabularx}
\usepackage{lipsum, amsmath, multirow}
\usepackage{siunitx, xcolor}
\usepackage{txfonts}
\usepackage[version=4]{mhchem}
\graphicspath{{figs/}{figsgaoerb/}} 
\usepackage[colorlinks,linkcolor=blue,anchorcolor=blue,citecolor=blue]{hyperref}

\begin{document}


\title{Non-Polynomial Wave Computation through Recurrent Resonant Scattering}


\author{Junyu Zhu}
\author{Enzong Wu}
\author{Xiaomeng Li}
\author{Hongsheng Chen}
\email{hansomchen@zju.edu.cn}
\author{Zuojia Wang}
\email{zuojiawang@zju.edu.cn}

\affiliation{Zhejiang Key Laboratory of Intelligent Electromagnetic Control and Advanced Electronic
Integration,  College of Information Science and Electronic Engineering, Zhejiang University, Hangzhou 310027, China.}



\date{\today}

\begin{abstract}
Structural nonlinearity enables nonlinear input--output mappings
to emerge from otherwise linear wave dynamics.
Repeated interactions with an input-encoded structure can enhance
such mappings, but finite-depth implementations restrict the
accessible functional order.
Here we show that recurrent scattering in a resonant cavity
provides a distinct regime for structural nonlinearity.
Repeated interactions are coherently accumulated into a
non-polynomial response to structural perturbations.
The resulting mapping naturally takes a Kolmogorov--Arnold form:
perturbation-induced resonance shifts realize the inner univariate
mappings, while the resonant spectral response provides the outer
mapping.
We identify two complementary physical controls of representation
capacity: the resonance linewidth governs the functional richness
within each branch, whereas combining multiple branches expands the
accessible function space.
Microwave-cavity measurements validate the two-stage mapping and
demonstrate a two-branch nonlinear computation through XOR
classification.
Our results connect recurrent resonant scattering with controllable
non-polynomial computation in linear wave systems.

\end{abstract}


\maketitle

\input{chapters/Introduction.tex}

\input{chapters/framework_2.tex}

\input{chapters/capacity.tex}

\input{chapters/experiment.tex}

\textit{Discussion.---}
Resonant feedback converts repeated coherent interactions into
a non-polynomial spectral response, allowing a linear wave system
to realize nonlinear mappings with respect to structural
perturbations.
Within the K--A representation, perturbation-induced resonance
shifts provide the inner mappings, while the resonant spectral
response supplies the outer functions.
The representation capacity is controlled at two complementary
levels: the resonance linewidth governs the functional richness
within each resonant branch, whereas additional branches expand
the network-level function space by combining multiple
compositional primitives.

We have demonstrated this framework numerically and experimentally.
The numerical results show that higher-order components of the
resonant response contribute directly to representation capacity,
while combining multiple branches further expands the accessible
function space.
The microwave experiment verifies the sum/difference-like inner
mappings and nonlinear outer mapping required for the
mirror-addressed two-branch XOR construction.
The present implementation operates in the weak-perturbation
regime, where resonance shifts remain approximately additive.
Extending the approach to independently addressable resonators
and larger branch numbers would enable more general physical
KAN architectures.

\begin{acknowledgments}
The work at Zhejiang University was sponsored by the National Natural Science Foundation of
China (NNSFC) (U25A20411,62222115), the Key Research and Development Program of Zhejiang
Province under Grant No.2024C01241(SD2).
\end{acknowledgments}

\bibliography{references.bib}

\end{document}


\setcounter{figure}{0}
\renewcommand{\thefigure}{S\arabic{figure}}

\onecolumngrid

\begin{center}
    {\LARGE\bfseries
    Supplemental Material for ``Non-Polynomial Wave Computation through
    Recurrent Resonant Scattering''\par}
    \vspace{1.2em}
\end{center}

\tableofcontents

\vspace{1.5em}
\twocolumngrid

\section{Lorentzian approximation and non-polynomial response}
\label{sec:supp_nonpolynomial}

We next give the assumptions leading to the Lorentzian response
used in the main text and clarify its non-polynomial character.

Near an isolated resonance, the transmission response can be
described by a dominant complex pole
\cite{fan_temporal_2003,kristensen_modeling_2020},
\begin{equation}
    S_{21}(\omega,\mathbf{x})
    \simeq
    S_{\mathrm{bg}}(\omega,\mathbf{x})
    +
    \frac{A(\mathbf{x})}
    {
        \omega-\widetilde{\omega}_{r}(\mathbf{x})
    } ,
    \label{eq:supp_s21}
\end{equation}
where
$S_{\mathrm{bg}}$
denotes the nonresonant background and
$A$
describes the excitation and readout coupling.

For the idealized response considered in the main-text derivation,
we neglect the nonresonant background and assume that the modal
coupling and linewidth vary weakly with the input,
\begin{equation}
    A(\mathbf{x})
    \simeq
    A_{0},
    \qquad
    \gamma(\mathbf{x})
    \simeq
    \gamma_{0}.
    \label{eq:supp_lorentzian_assumptions}
\end{equation}
Writing the input-dependent resonance frequency as
\begin{equation}
    \omega_{r}(\mathbf{x})
    =
    \omega_{0}
    +
    u(\mathbf{x}),
    \label{eq:supp_shifted_frequency}
\end{equation}
the resonant transmission intensity at a fixed readout frequency
$\omega_m$ becomes
\begin{equation}
    L_m(u)
    =
    \frac{B_m}
    {
        (\Delta_m-u)^2
        +
        \gamma_{0}^{2}
    },
    \label{eq:supp_lorentzian}
\end{equation}
where
\begin{equation}
    \Delta_m
    =
    \omega_m-\omega_0,
    \qquad
    B_m
    =
    |A_0|^2 .
    \label{eq:supp_detuning}
\end{equation}

The response in
Eq.~(\ref{eq:supp_lorentzian})
is a rational function of $u$ and therefore cannot, in general,
be represented by a finite-order polynomial.
Its local higher-order content can be seen by writing
\begin{equation}
    L_m(u)
    =
    \frac{B_m}{2i\gamma_0}
    \left[
        \frac{1}
        {\Delta_m-u-i\gamma_0}
        -
        \frac{1}
        {\Delta_m-u+i\gamma_0}
    \right].
    \label{eq:supp_partial_fraction}
\end{equation}
For
\begin{equation}
    |u|
    <
    \sqrt{
        \Delta_m^2+\gamma_0^2
    },
    \label{eq:supp_convergence}
\end{equation}
each denominator admits a geometric expansion,
\begin{equation}
    \frac{1}
    {\Delta_m\mp i\gamma_0-u}
    =
    \frac{1}
    {\Delta_m\mp i\gamma_0}
    \sum_{n=0}^{\infty}
    \left(
        \frac{u}
        {\Delta_m\mp i\gamma_0}
    \right)^n .
    \label{eq:supp_lorentzian_expansion}
\end{equation}
Consequently,
\begin{equation}
    L_m(u)
    =
    \sum_{n=0}^{\infty}
    c_{m,n}u^n .
    \label{eq:supp_infinite_series}
\end{equation}
For a generic detuning,
the expansion contains terms of arbitrarily high order in $u$.
At exact resonance, symmetry removes the odd-order terms,
while arbitrarily high even-order terms remain.

The series in
Eq.~(\ref{eq:supp_infinite_series})
is only a local representation of the response.
The physical mapping is the Lorentzian rational function in
Eq.~(\ref{eq:supp_lorentzian}), which is not restricted to any
finite polynomial order.

\section{Cavity perturbation and spatial inner mapping}

The spatial inner mapping used in the main text is obtained from
first-order cavity perturbation theory. The cavity is modeled as an
air-filled rectangular resonator with dimensions
$22.86\times19.87\times12~\mathrm{mm}^3$. The lowest resonant mode
used here is the $\mathrm{TM}_{110}$ mode, whose ideal resonance
frequency is
\begin{equation}
f_{110}
=
\frac{c}{2}
\sqrt{
\left(\frac{1}{L_x}\right)^2+
\left(\frac{1}{L_y}\right)^2
}
\simeq 10~\mathrm{GHz}.
\end{equation}

For a weak localized perturbation, Slater's cavity-perturbation
theorem gives the first-order resonance shift in the form
\cite{doi:10.1049/pi-c.1960.0041}
\begin{equation}
\Delta\omega
=
C
\int_{\Omega_p}
g(\mathbf r)\,dV ,
\qquad
g(\mathbf r)
=
\mu_0|\mathbf H_0(\mathbf r)|^2
-
\epsilon_0|\mathbf E_0(\mathbf r)|^2 ,
\label{eq:sm_slater}
\end{equation}
where $\mathbf E_0$ and $\mathbf H_0$ are the fields of the
unperturbed cavity mode, $\Omega_p$ is the perturbation volume, and
$C$ contains the modal normalization and perturbation-dependent
prefactor.

For a sufficiently small perturbing object, the field variation
across its volume is weak, so that
\begin{equation}
\Delta\omega(\mathbf r_p)
\simeq
C V_p g(\mathbf r_p),
\end{equation}
where $\mathbf r_p$ denotes the perturbation position. Thus, apart
from an overall scale factor, the spatial resonance-shift landscape
is determined by $g(\mathbf r)$. In Fig.~\ref{fig:supp_background_basis}(a) of the main text we
therefore use the normalized quantity
\begin{equation}
g_n(\mathbf r)
=
\frac{g(\mathbf r)}
{\max_{\mathbf r}|g(\mathbf r)|}.
\end{equation}

When the $i$th perturbation moves along a prescribed trajectory
$\mathbf r_i(x_i)$, its resonance shift defines the univariate inner
mapping
\begin{equation}
\phi_i(x_i)
\equiv
\Delta\omega_i[\mathbf r_i(x_i)].
\end{equation}
For multiple weak perturbations, the first-order shifts are
approximately additive,
\begin{equation}
\Delta\omega(x_1,\ldots,x_N)
\simeq
\sum_i \phi_i(x_i),
\end{equation}
which gives the additive inner mapping used in the
Kolmogorov--Arnold representation of the main text.

\section{Reshaping of resonant basis functions}
\label{sec:supp_linewidth_background}

The idealized model in the main text assumes that the input primarily
shifts the resonance frequency, while the other parameters of the
single-pole response remain approximately unchanged. More generally,
at a fixed readout frequency $\omega_m$, the transmission can be written
as
\begin{equation}
    S_{21}(\omega_m,x)
    =
    S_{\mathrm{bg},m}(x)
    +
    \frac{A(x)}
    {\Delta_m-u(x)+i\gamma(x)},
    \label{eq:supp_general_s21}
\end{equation}
where
\begin{equation}
    \Delta_m=\omega_m-\omega_0,
    \qquad
    S_{\mathrm{bg},m}(x)
    \equiv S_{\mathrm{bg}}(\omega_m,x).
\end{equation}
The corresponding fixed-frequency physical basis is
\begin{equation}
    L_m(x)
    =
    \left|
        S_{21}(\omega_m,x)
    \right|^2.
    \label{eq:supp_general_basis}
\end{equation}
The Lorentzian basis used in the main-text derivation is recovered when
\begin{equation}
    A(x)=A_0,
    \qquad
    \gamma(x)=\gamma_0,
    \qquad
    S_{\mathrm{bg},m}(x)=0.
    \label{eq:supp_ideal_parameters}
\end{equation}
Thus, input dependence of $A$, $\gamma$, or $S_{\mathrm{bg}}$
primarily appears as a reshaping of the fixed-frequency physical
basis functions.
The present numerical study keeps $A$ fixed and uses linewidth and coherent-background
variations as two representative sources of basis reshaping.
This input-dependent linewidth variation should be distinguished
from the uniform linewidth, or equivalently the global $Q$,
considered in the main-text capacity analysis.
There, the linewidth sets the overall width and coverage of the
entire basis bank, whereas here $\gamma(x)$ locally deforms an
otherwise fixed spectral basis.

We consider a dimensionless, noiseless model with
\begin{equation}
    x\in[-1,1],
    \qquad
    u(x)=3x,
    \qquad
    A=1,
    \qquad
    \gamma_0=1.
    \label{eq:supp_synthetic_model}
\end{equation}
Fifteen readout detunings are uniformly distributed over
$\Delta_m\in[-3,3]$. The linewidth and coherent background are
parameterized as
\begin{align}
    \gamma(x)
    &=
    1+\alpha_\gamma g(x),
    \label{eq:supp_linewidth_variation}
    \\
    S_{\mathrm{bg},m}(x)
    &=
    \beta h(x),
    \label{eq:supp_background_variation}
\end{align}
with
\begin{equation}
    \max_x |g(x)|=1,
    \qquad
    \max_x |h(x)|=1,
\end{equation}
where $g(x)$ is real and $h(x)$ is complex. Random profiles are generated
from Fourier series through the third harmonic. We scan
$\alpha_\gamma=0$--$0.30$ and $\beta=0$--$0.50$, with 100 paired random
realizations at each parameter point.

\begin{figure}[t]
    \centering
    \includegraphics[width=\linewidth]
    {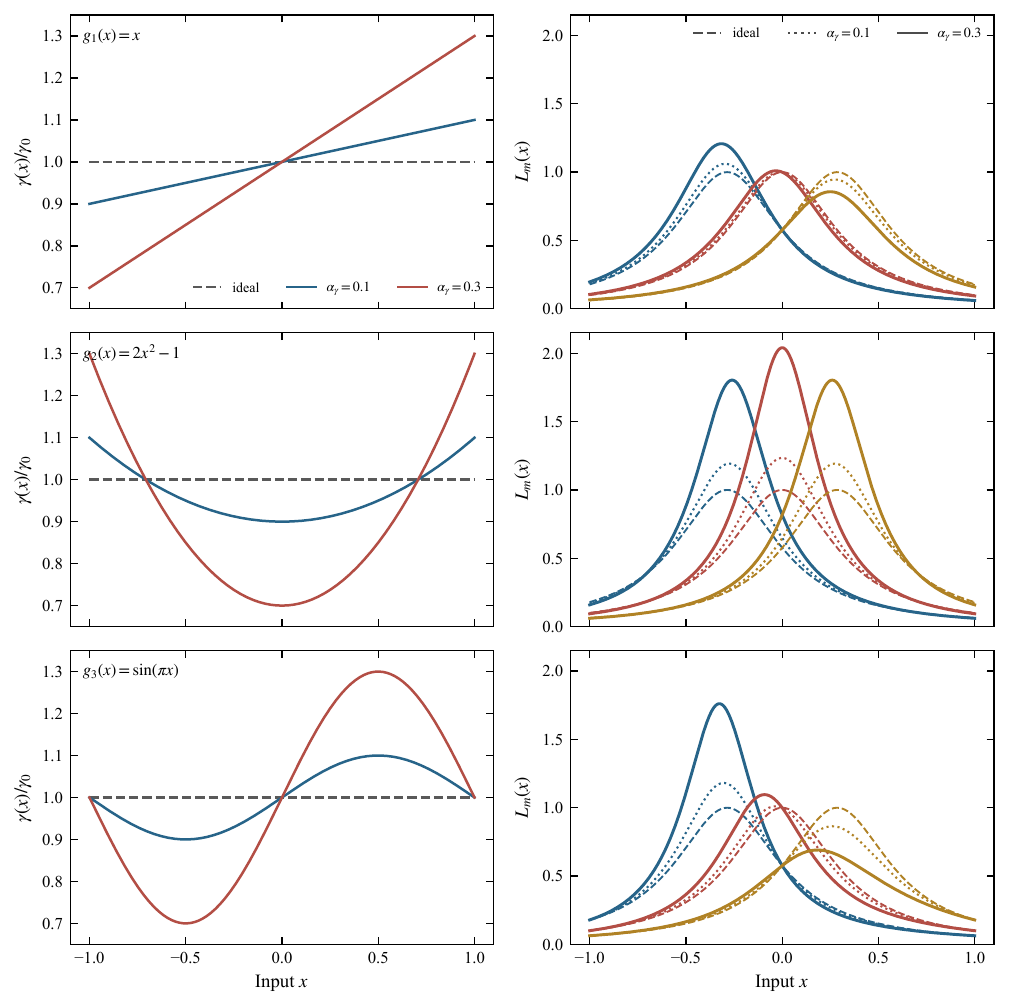}
    \caption{Representative linewidth profiles and the corresponding
    fixed-frequency basis functions for $\Delta_m=-6/7$, $0$, and $+6/7$
    (blue, red, and ochre, respectively).}
    \label{fig:supp_linewidth_basis}
\end{figure}

Input-dependent linewidth changes the local peak height, width, and
curvature of $L_m(x)$. Because $\gamma$ itself varies along the input
coordinate, it can also introduce asymmetry and an apparent displacement
of the basis maximum even when the resonance-shift mapping $u(x)$ is
unchanged. Figure~\ref{fig:supp_linewidth_basis} illustrates this effect
for three representative profiles,
$g(x)=x$, $2x^2-1$, and $\sin(\pi x)$.

\begin{figure}[t]
    \centering
    \includegraphics[width=\linewidth]
    {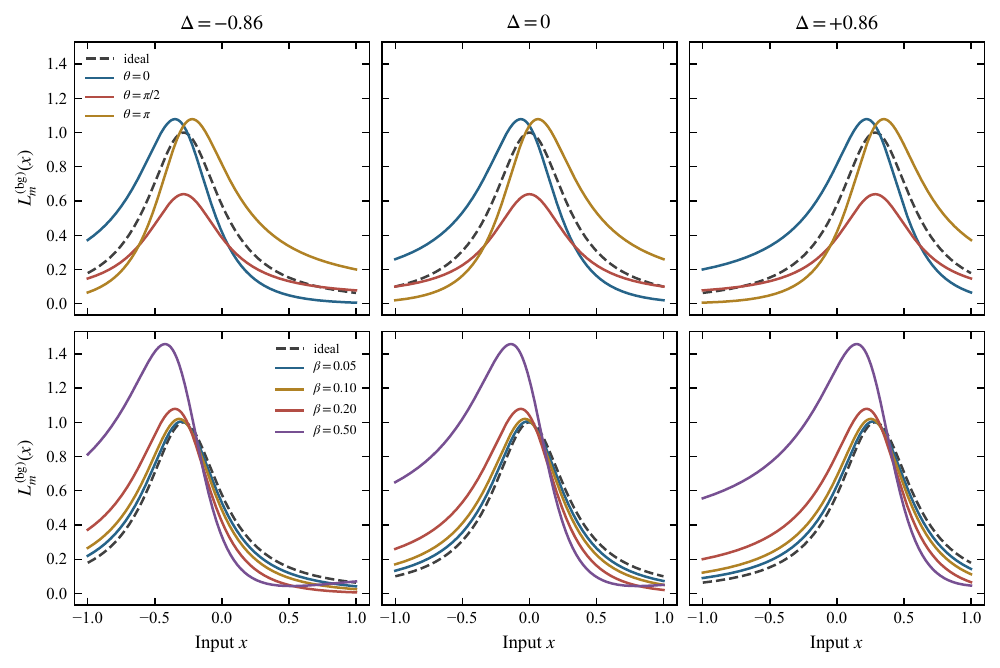}
    \caption{Reshaping of the resonant basis by a constant coherent
    background. The upper row varies the background phase at
    $\beta=0.2$; the lower row varies its strength at $\theta=0$.}
    \label{fig:supp_background_basis}
\end{figure}

A coherent background reshapes the basis through field-level
interference. For a constant background
\begin{equation}
    S_{\mathrm{bg}}
    =
    \beta e^{i\theta},
    \qquad
    d=\Delta_m-3x,
    \label{eq:supp_constant_background}
\end{equation}
and $\gamma=1$, Eq.~\eqref{eq:supp_general_basis} becomes
\begin{align}
    L_m(d)
    &=
    \beta^2
    +
    \frac{1-2\beta\sin\theta}
    {d^2+1}
    \nonumber\\
    &\quad+
    \frac{2\beta d\cos\theta}
    {d^2+1}.
    \label{eq:supp_coherent_interference}
\end{align}
The last term is odd about resonance and produces a Fano-like
asymmetry \cite{fan_temporal_2003}.
Consequently, $\theta=0$ and $\theta=\pi$ generate opposite skewing,
whereas a background in quadrature, $\theta=\pi/2$, suppresses the
symmetric resonant contribution. Representative phase and amplitude
variations are shown in Fig.~\ref{fig:supp_background_basis}.

Once the physical basis changes, two effects should be distinguished.
First, weights trained for the ideal basis are no longer matched to the
reshaped basis, producing a calibration mismatch. Second, the function space
spanned by the reshaped basis can itself differ from that of the ideal
bank. To separate these effects, let $B_0$ and $B$ denote the ideal and
reshaped basis matrices. We first obtain
\begin{equation}
    \mathbf{w}_0
    =
    \underset{\mathbf{w}}{\operatorname{argmin}}
    \left\|
        \mathbf{f}-B_0\mathbf{w}
    \right\|_2^2,
    \label{eq:supp_ideal_weights}
\end{equation}
and then compare the fixed-weight prediction $B\mathbf{w}_0$ with the
recalibrated solution
\begin{equation}
    \mathbf{w}^{\ast}
    =
    \underset{\mathbf{w}}{\operatorname{argmin}}
    \left\|
        \mathbf{f}-B\mathbf{w}
    \right\|_2^2.
    \label{eq:supp_retrained_weights}
\end{equation}
Recalibration therefore changes only the linear readout weights; it does
not restore the ideal basis functions or modify the readout frequencies.

\begin{figure}[t]
    \centering
    \includegraphics[width=\linewidth]
    {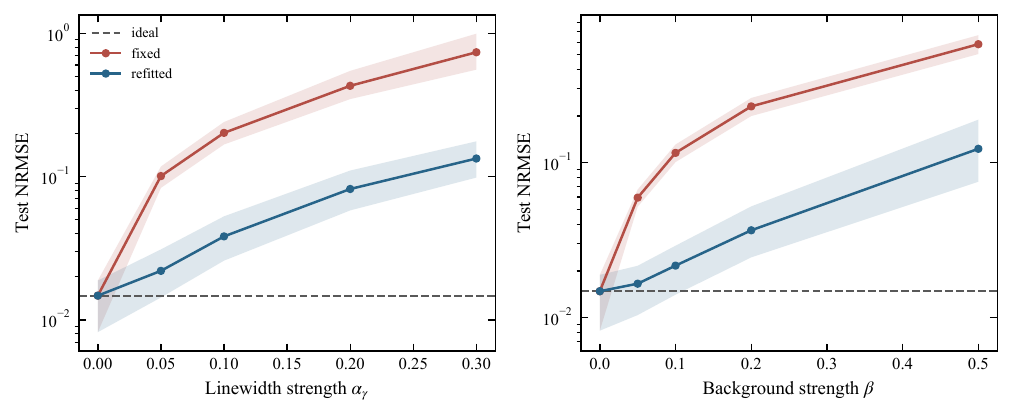}
    \caption{Test NRMSE under separate random linewidth and
    coherent-background variations. Curves and shaded regions show the
    median and interquartile range over 100 random profiles and 20 shared
    target functions.}
    \label{fig:supp_calibration}
\end{figure}

We evaluate these two cases using the same 20 fixed $K=4$ random Fourier
targets for all perturbation conditions. The ideal 15-function basis bank
has a median test NRMSE of $0.0148$, with an interquartile range of
$0.0082$--$0.0188$. For random linewidth variation,
$\alpha_\gamma=0.2$ increases the median NRMSE to $0.430$ with fixed
ideal weights, while recalibration reduces it to $0.0820$. At
$\alpha_\gamma=0.3$, the corresponding values are $0.738$ and $0.134$.
For random coherent backgrounds, $\beta=0.2$ gives $0.231$ and $0.0366$,
whereas $\beta=0.5$ gives $0.582$ and $0.123$. These trends are summarized
in Fig.~\ref{fig:supp_calibration}.

\begin{figure}[t]
    \centering
    \includegraphics[width=\linewidth]
    {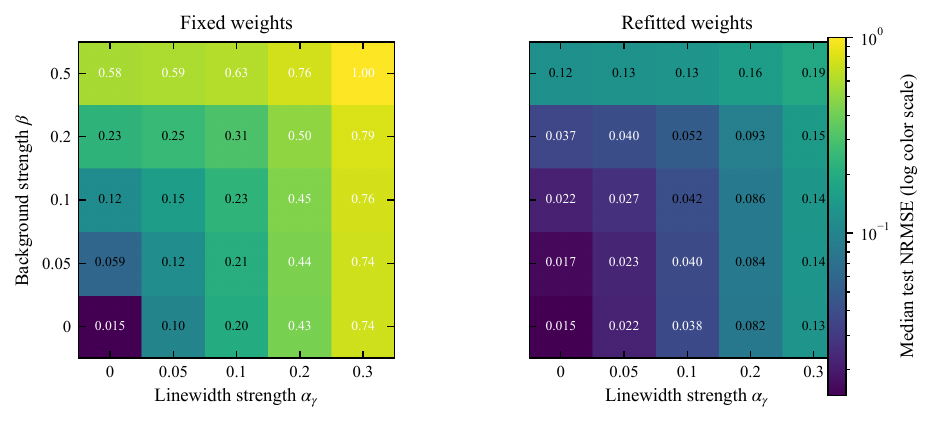}
    \caption{Median test NRMSE for simultaneous linewidth and
    coherent-background variations using fixed ideal weights (left) and
    recalibrated weights (right). Both maps use the same logarithmic color
    scale.}
    \label{fig:supp_joint_robustness}
\end{figure}

The same distinction persists when both effects are present. At
$(\alpha_\gamma,\beta)=(0.2,0.2)$, the median NRMSE is $0.500$ with the
fixed ideal weights and $0.0930$ after recalibration. At the strongest
condition considered, $(0.3,0.5)$, the corresponding values are $1.003$
and $0.190$ [Fig.~\ref{fig:supp_joint_robustness}]. Recalibration therefore
removes a large fraction of the error caused by weight mismatch. However,
the remaining error is still well above the ideal value of $0.0148$,
showing that basis reshaping also changes the accessible function space.

\begin{figure}[t]
    \centering
    \includegraphics[width=0.78\linewidth]
    {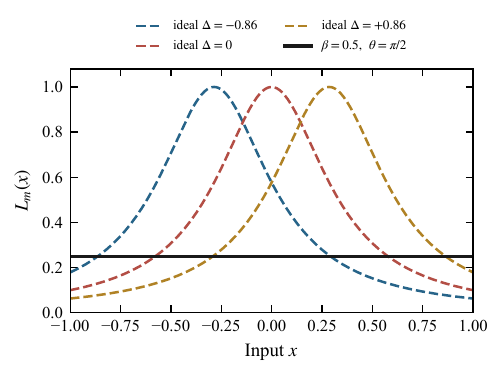}
    \caption{Exact coherent cancellation at $\beta=0.5$ and
    $\theta=\pi/2$. All 15 physical basis functions collapse to the same
    constant response.}
    \label{fig:supp_basis_collapse}
\end{figure}

Recalibration cannot recover representation capacity when coherent
interference makes the basis bank itself degenerate. From
Eq.~\eqref{eq:supp_coherent_interference}, the special case
\begin{equation}
    \theta=\frac{\pi}{2},
    \qquad
    \beta=0.5
\end{equation}
removes the resonant term and gives
\begin{equation}
    L_m(x)=0.25
\end{equation}
for every input and all 15 readout channels. The basis matrix then has
rank one, and refitting the weights cannot represent the nonconstant,
zero-mean Fourier targets; the median NRMSE is approximately $1.000$.
This exact cancellation is shown in
Fig.~\ref{fig:supp_basis_collapse}.

\begin{figure}[t]
    \centering
    \includegraphics[width=\linewidth]
    {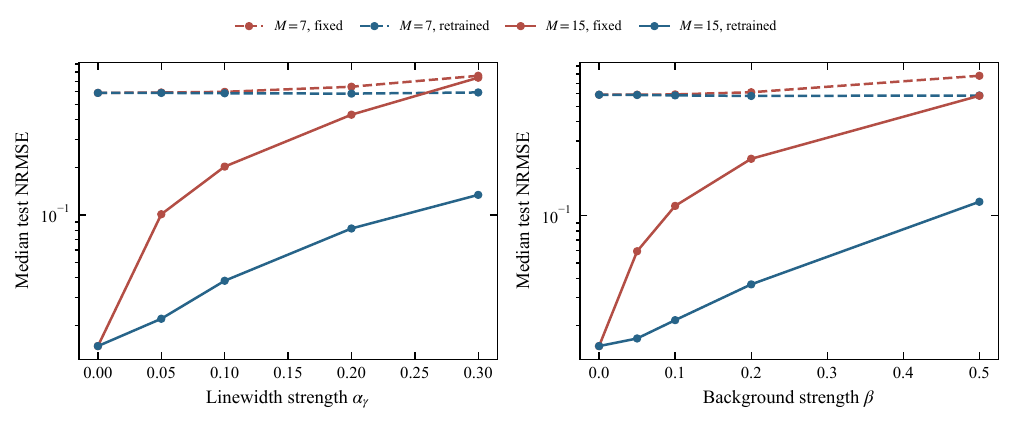}
    \caption{Comparison of seven- and fifteen-function basis banks over
    the same detuning range, target ensemble, and perturbation profiles.
    The higher-capacity 15-function bank makes the residual representation
    loss after basis reshaping more visible.}
    \label{fig:supp_basis_number}
\end{figure}

The effect of basis reshaping becomes substantially clearer when the ideal
basis bank itself has high approximation accuracy. Repeating the same
analysis with only seven readout frequencies over the same detuning range
gives an ideal median NRMSE of $0.591$, compared with $0.0148$ for
$M=15$. 
The large approximation floor of the seven-function bank makes
the additional loss caused by basis deformation much less visible.
In contrast, the lower ideal error of the 15-function bank exposes the residual penalty
after recalibration, as shown in Fig.~\ref{fig:supp_basis_number}.
Importantly, the 15-function bank still gives lower \emph{absolute}
recalibrated errors than the seven-function bank under the same
perturbations; it is simply more sensitive relative to its much lower
ideal baseline.

The denser 15-function bank also contains more strongly correlated
neighboring basis functions. After column-RMS normalization, the condition
number of the ideal basis matrix increases from approximately $11.7$ for
$M=7$ to $681.5$ for $M=15$. This does not by itself explain the residual
error in the present noiseless least-squares calculation, but it indicates
that the high-accuracy representation relies on more finely coordinated
linear combinations of neighboring basis functions. Strong basis
reshaping can therefore modify this high-accuracy function space even when
the readout weights are refitted.

Overall, input-dependent linewidth and coherent background produce both a
calibration mismatch and an intrinsic deformation of the resonant basis
space. Recalibrating the physical basis and retraining the linear readout
weights removes much of the former, but it does not in general restore the
ideal function space. This residual effect is most visible when the ideal
basis bank has sufficiently high approximation capacity. A practical
implementation should therefore train on calibrated physical basis
functions while also maintaining a stable and nondegenerate basis bank.

\section{Numerical methods for representation-capacity analysis}

\subsection{Single resonant response and Chebyshev expansion}

Fig.~3(a) and 3(b) characterize the functional content of a
single resonant response. We use
\begin{equation}
S(x)
=
\operatorname{Re}
\frac{1}{\gamma+i(\Delta-\alpha x)}
=
\frac{\gamma}
{\gamma^2+(\Delta-\alpha x)^2},
\end{equation}
with $\omega_0=1$, $\Delta=0.023$, $\alpha=0.1$, and
$x\in[-1,1]$. The linewidth is
\begin{equation}
\gamma=\frac{\omega_0}{2Q},
\end{equation}
and $Q=5$, $10$, and $20$ are compared. No amplitude
normalization is applied in Fig.~3(a) and 3(b).

For Fig.~3(b), the response is expanded in Chebyshev polynomials,
\begin{equation}
S(x)=\sum_{n=0}^{\infty}c_nT_n(x).
\end{equation}
The coefficients are evaluated using $J=1024$
Gauss--Chebyshev nodes,
\begin{equation}
\theta_j=\frac{\pi(j+1/2)}{J},
\qquad
x_j=\cos\theta_j ,
\end{equation}
with
\begin{align}
c_0 &=
\frac{1}{J}\sum_{j=0}^{J-1}S(x_j),\\
c_n &=
\frac{2}{J}\sum_{j=0}^{J-1}
S(x_j)\cos(n\theta_j),
\qquad n\geq1 .
\end{align}
Coefficients are evaluated up to $n=256$, while Fig.3(b)
displays $|c_n|$ up to $n=120$.

\subsection{Single-branch approximation of random Fourier targets}

For Fig.3(c), the input produces a linear resonance shift,
\begin{equation}
u(x)=\alpha x,
\qquad
\frac{\omega_r}{\omega_0}=1+0.1x .
\end{equation}
The response is sampled at $M=15$ fixed readout frequencies,
giving the basis functions
\begin{equation}
L_m(x)
=
\frac{\gamma}
{\gamma^2+[\Delta_m-\alpha x]^2},
\qquad
\gamma=\frac{\omega_0}{2Q}.
\end{equation}
The detunings $\Delta_m$ are uniformly distributed over
$[-0.12,0.12]\omega_0$, corresponding to readout frequencies from
$0.88\omega_0$ to $1.12\omega_0$. The same frequency set is used
for $Q=5$, $10$, and $20$.
To construct the finite-order reference used in Fig.3(c),
we take the $Q=10$ basis bank and expand each resonant basis
function in Chebyshev polynomials,
\begin{equation}
L_m(x)
=
\sum_{n=0}^{\infty} c_{m,n} T_n(x).
\end{equation}
The reference basis is obtained by retaining only terms through
fifth order,
\begin{equation}
L_m^{(5)}(x)
=
\sum_{n=0}^{5} c_{m,n} T_n(x).
\end{equation}
Thus, the full and truncated models use the same 15 readout
channels and the same $Q=10$ resonant responses as their starting
point, while the truncated bank contains no Chebyshev components
above $T_5$.

The target functions are random Fourier series,
\begin{equation}
f_K(x)
=
\frac{
\displaystyle
\sum_{k=1}^{K}
\left[
a_k\cos(k\pi x)
+
b_k\sin(k\pi x)
\right]
}{
\displaystyle
\sqrt{
\frac{1}{2}
\sum_{k=1}^{K}
(a_k^2+b_k^2)
}
},
\qquad K=1,\ldots,6 ,
\end{equation}
where the Fourier coefficients are independently drawn from a
standard normal distribution,
$a_k,b_k\sim\mathcal{N}(0,1)$.
The normalization gives each target unit RMS over $[-1,1]$.

For each $K$, 20 target functions are generated using the random
seed 20260913. Targets of different complexity are constructed from
the same coefficient realizations by retaining the first $K$
harmonics and renormalizing. The same target ensemble is used for
all values of $Q$.

The prediction is
\begin{equation}
\hat f(x)
=
\sum_{m=1}^{15}w_mL_m(x).
\end{equation}
The weights are fitted on 401 uniformly spaced training points.
Before fitting, each basis function is normalized by its training
RMS,
\begin{equation}
Z_m(x)=\frac{L_m(x)}{r_m},
\qquad
r_m=
\sqrt{
\left\langle L_m^2(x)\right\rangle_{\rm train}
}.
\end{equation}
The normalized weights $a_m$ are obtained from
\begin{equation}
\min_{\{a_m\}}
\left[
\frac{1}{N_{\rm tr}}
\sum_j
\left(
\sum_m a_mZ_m(x_j)-f_K(x_j)
\right)^2
+
\lambda\sum_m a_m^2
\right],
\end{equation}
with $\lambda=10^{-10}$ and no output bias.
The physical weights are $w_m=a_m/r_m$.

The fitted weights are evaluated on 4096 uniformly sampled test
points. The test error is
\begin{equation}
\mathrm{NRMSE}
=
\frac{
\sqrt{
\left\langle
[\hat f(x)-f_K(x)]^2
\right\rangle_{\rm test}
}
}{
\operatorname{std}_{\rm test}[f_K(x)]
}.
\end{equation}
Fig.~3(c) shows the median and interquartile range over the
20 target functions.

\subsection{Multi-branch benchmark}

For Fig.~3(d), both models approximate
\begin{equation}
f(x,y)
=
\frac{x+y}{1+xy},
\qquad
(x,y)\in[-0.8,0.8]^2 ,
\end{equation}
using the topology $[2,B,1]$ with
\begin{equation}
B\in\{1,2,3,4,6,8\}.
\end{equation}
The training and validation sets contain 4000 and 1000 uniformly
sampled random points, respectively, and testing is performed on a
$201\times201$ uniform grid. For each $B$, ten independent
initializations are trained using the same data.

For the resonant KAN, each branch contains two independently
trainable perturbation trajectories,
\begin{equation}
u_q(x,y)
=
\beta g[\mathbf r_{q1}(x)]
+
\beta g[\mathbf r_{q2}(y)],
\end{equation}
with
\begin{equation}
g(X,Y)
=
\sin^2(\pi X)\sin^2(\pi Y).
\end{equation}
Here $g(X,Y)$ denotes a dimensionless benchmark landscape used in
the numerical capacity analysis.
Each trajectory is represented by a cubic clamped B spline with
five two-dimensional control points. 
The resonant parameters are
\begin{equation}
Q=1000,
\qquad
\gamma=5\times10^{-4},
\qquad
\beta=7.5\times10^{-4}.
\end{equation}
Each branch uses 15 fixed spectral channels with detunings
uniformly distributed over $\Delta_m\in[0,2\beta]$.
The unit-peak resonant features are
\begin{equation}
A_{qm}
=
\frac{1}{
1+[(\Delta_m-u_q)/\gamma]^2
},
\end{equation}
and the network output is
\begin{equation}
\hat f(x,y)
=
b+
\sum_{q=1}^{B}
\sum_{m=1}^{15}
a_{qm}A_{qm}(x,y).
\end{equation}
The linear readout weights and bias are solved by variable
projection for the current trajectories. The optimized objective is
\begin{equation}
\mathcal L
=
\mathrm{MSE}
+
10^{-12}\mathcal R_{\rm smooth}
+
10^{-10}\|\mathbf a\|_2^2 ,
\end{equation}
where $\mathcal R_{\rm smooth}$ penalizes the second differences of
the trajectory control points.

The digital baseline uses the standard \texttt{pykan}
implementation \cite{liu_kan_2025} with width $[2,B,1]$,
cubic spline edges (\texttt{grid}=10, \texttt{k}=3), and the
default SiLU base function.
Symbolic processing and pruning are disabled.

Both models are trained using 300 full-batch Adam steps followed by
120 L-BFGS outer calls. The Adam learning rates are $0.025$ for the
resonant model and $0.01$ for the digital KAN. The model state with
the lowest validation MSE among the recorded states is retained.

The test NRMSE is defined as
\begin{equation}
\mathrm{NRMSE}
=
\frac{
\sqrt{
N_{\rm test}^{-1}
\sum_k
[\hat f_k-f_k]^2
}
}{
\operatorname{std}(f_{\rm test})
}.
\end{equation}
Figure~3(d) shows the median and interquartile range over ten
independent initializations.

The comparison is branch matched rather than strictly
parameter matched. Including the linear readout, the resonant model
contains $35B+1$ trainable parameters, whereas the digital KAN
contains approximately $45B$ trainable parameters.

\section{Experimental setup and data analysis}

\subsection{Microwave cavity and experimental setup}

The experimental platform consists of a CNC-machined copper
rectangular microwave cavity with internal dimensions
\begin{equation}
    L_x\times L_y\times L_z
    =
    22.86\times19.87\times12~\mathrm{mm}^3 .
\end{equation}

The cavity is excited and measured through two SMA ports.
Each port is coupled to the cavity through a cylindrical probe
with radius $0.25$~mm and length $6$~mm. The cylindrical air
region surrounding the probe has a radius of $2.05$~mm.
The measured quantity is the transmission coefficient $S_{21}$.

Two magnetic spheres with a diameter of $3$~mm are used as
movable structural perturbations. The spheres are located inside
the cavity and are actuated magnetically by external magnets
mounted on two independently controlled translation stages.
The magnetic actuation therefore allows the perturbation positions
to be varied without introducing a mechanical feedthrough into the
cavity.

The two perturbation trajectories are arranged approximately
symmetrically with respect to the cavity center line. For each
perturbation, the local trajectory extends from
$(u,v)=(1.0,5.0)$~mm to $(8.6,5.0)$~mm with a spacing of
$0.4$~mm, giving 20 discrete positions. Joint scanning of the
two perturbations therefore produces a $20\times20$ grid containing
400 position combinations.
After sorting the physical trajectory coordinates, the two inputs
are normalized as
\begin{equation}
x_i=-1+\frac{2i}{19},
\qquad
y_j=-1+\frac{2j}{19},
\qquad i,j=0,\ldots,19.
\end{equation}

\subsection{Scattering measurement}

At each pair of perturbation positions, the transmission spectrum
$S_{21}(f)$ is measured over
\begin{equation}
    9.5~\mathrm{GHz}
    \leq f \leq
    10.5~\mathrm{GHz}
\end{equation}
using 501 uniformly spaced frequency points, corresponding to a
frequency interval of $2$~MHz.

The measurement is performed using a Ceyear 3672C vector network
analyzer. After both translation stages reach their prescribed
positions, a waiting time of $0.5$~s is applied before the spectrum
is acquired.

\begin{figure}[t]
    \centering
    \includegraphics[width=\columnwidth]{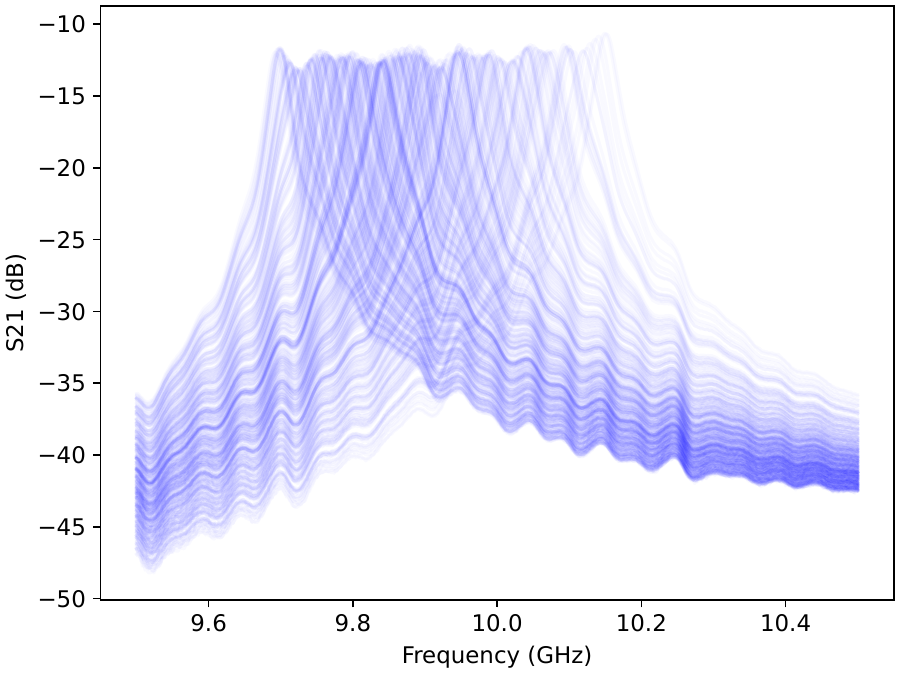}
    \caption{
    Measured transmission spectra for all 400 combinations of the
    two perturbation positions. The displacement of the dominant
    resonance reflects the input-dependent cavity-frequency shift.
    }
    \label{fig:all_spectra}
\end{figure}

All 400 position combinations are successfully measured using the
same frequency grid, with no missing or duplicated spectra.
The complete set of measured transmission spectra is shown in
Fig.~\ref{fig:all_spectra}. The dominant resonance shifts continuously across the
measurement grid while remaining within the selected frequency
window.
The measured $S_{21}$ values are stored in decibels and converted
to linear power according to
\begin{equation}
    P(x_i,y_j;f_k)
    =
    10^{S_{21,\mathrm{dB}}(x_i,y_j;f_k)/10}.
    \label{eq:sm_power_conversion}
\end{equation}

\subsection{Extraction of the resonance-frequency map}

The resonance frequency is extracted from each measured transmission
spectrum for physical characterization of the perturbation-induced
inner mapping. The resonance frequency is not used as an input to
the classification or regression models described below.

For each spectrum, the measured $S_{21}$ values are first converted
to linear power using Eq.~(\ref{eq:sm_power_conversion}). The power
spectrum is then smoothed using a Gaussian kernel with a standard
deviation of one frequency sample, corresponding to $2$~MHz.

The largest peak of the smoothed spectrum is identified at frequency
$f_k$. To reduce discretization of the peak position by the
$2$-MHz sampling interval, a three-point parabolic interpolation is
applied to the peak and its two neighboring samples. The fractional
frequency offset is
\begin{equation}
    \delta
    =
    \frac{
        P_{k-1}-P_{k+1}
    }{
        2(P_{k-1}-2P_k+P_{k+1})
    },
    \label{eq:sm_peak_interpolation}
\end{equation}
and the interpolated resonance frequency is
\begin{equation}
    f_r
    =
    f_k+\delta\Delta f ,
\end{equation}
where $\Delta f=2$~MHz. The interpolation offset is restricted to
$|\delta|\leq0.5$.

Across the complete $20\times20$ measurement grid, the extracted
resonance frequencies range from
$9.69238$ to $10.15194$~GHz. No extracted resonance occurs at the
boundary of the measured frequency interval, and no competing peak
exceeding 80\% of the principal peak amplitude is found according
to the peak-selection criterion used in the analysis.

The interpolation procedure provides a sub-grid estimate of the
spectral peak position and should not be interpreted as implying a
measurement accuracy equal to the numerical interpolation
resolution.

\subsection{Additivity of the measured resonance shift}

The perturbative model in the main text assumes that the frequency
shifts induced by the two encoded inputs are approximately additive.
We test this assumption directly using the measured two-dimensional
resonance-frequency map.

The measured map is decomposed as
\begin{equation}
    f_r(x_i,y_j)
    =
    f_0+p_i+q_j+\epsilon_{ij},
    \label{eq:sm_additive_decomposition}
\end{equation}
where
\begin{equation}
f_0=\langle f_r\rangle_{i,j},
\qquad
p_i=\langle f_r(x_i,y_j)\rangle_j-f_0,
\qquad
q_j=\langle f_r(x_i,y_j)\rangle_i-f_0,
\end{equation}
and $\epsilon_{ij}$ is the residual interaction term.

The interaction residual ratio is defined as
\begin{equation}
\eta_{\mathrm{int}}
=
\frac{\|\epsilon\|_F}
{\|f_r-f_0\|_F},
\end{equation}
where $\|\cdot\|_F$ denotes the Frobenius norm. For the measured
resonance-frequency map, we obtain
$\eta_{\mathrm{int}}=0.130$, with a residual RMS of
$14.29$~MHz.

The measured resonance landscape is therefore predominantly
described by two additive one-dimensional contributions, while
finite deviations from ideal additivity remain. These deviations can
originate from higher-order perturbation effects, mutual interaction
between the two perturbing elements, and changes in the cavity mode
profile as the perturbations move.

The experimentally measured responses, rather than the ideal
additive approximation, are used directly for all subsequent
classification and regression tasks. The computational results
therefore do not require exact validity of
Eq.~(\ref{eq:sm_additive_decomposition}).

\subsection{Construction of the two resonant branches}

The ideal XOR construction considered in the main text requires two
branches associated with sum- and difference-like inner coordinates.
In the present proof-of-principle experiment, only one physical
cavity is measured. The second branch is constructed by reversing
the encoding direction of the $y$ input.

We define the directly measured branch as
\begin{equation}
    P^{(+)}(x_i,y_j;f_k)
    =
    P(x_i,y_j;f_k),
\end{equation}
and the mirror-addressed branch as
\begin{equation}
    P^{(-)}(x_i,y_j;f_k)
    =
    P(x_i,y_{19-j};f_k)
    =
    P(x_i,-y_j;f_k).
    \label{eq:sm_mirror_branch}
\end{equation}

Because the measured grid contains both members of every mirrored
$y$ pair, the second branch is constructed entirely from measured
spectra and requires no interpolation.

The binary XOR target is
\begin{equation}
    t(x,y)
    =
    \mathbf{1}(xy<0),
\end{equation}
while the corresponding continuous regression target is
\begin{equation}
    g(x,y)=-xy.
    \label{eq:sm_product_target}
\end{equation}

For the ideal quadratic construction,
\begin{equation}
    \frac{(x-y)^2-(x+y)^2}{4}
    =
    -xy.
\end{equation}
The reversed subtraction order is therefore chosen such that the
shared two-branch output has the same sign as the continuous target
$-xy$.

\subsection{Selection of fixed-frequency readout channels}

The computational readout uses the measured power at a discrete set
of fixed frequencies. We consider
\begin{equation}
    N_b\in
    \{1,2,3,5,7,10,13,16\}
\end{equation}
readout channels.

For every training set, the frequency channels are selected using
only the first-branch spectra contained in that training set. At
each measured frequency $f_k$, we calculate the power variance
across the training positions,
\begin{equation}
    V_k
    =
    \operatorname{Var}_{\mathrm{train}}
    \left[
        P^{(+)}(x,y;f_k)
    \right].
\end{equation}

The effective spectral interval is defined by the first and last
frequency samples satisfying
\begin{equation}
    V_k
    \geq
    0.1\,\max_{k}V_k.
\end{equation}

The requested number $N_b$ of readout channels is then distributed
uniformly over this interval and rounded to the nearest measured
frequency sample. For $N_b=1$, the central frequency of the
effective interval is used.

The same selected frequencies are used for both resonant branches
and for all resonant baseline models within a given training fold.
Frequency selection is repeated independently inside every outer
and inner cross-validation training set and does not use test labels.

For reference, when the model is retrained on the complete measured
grid, the five selected readout frequencies are
\[
\begin{split}
f_k =\{&
9.684,\,
9.804,\,
9.922,\,
10.040,\,
10.160
\}\ {\rm GHz},
\end{split}
\]
For the sixteen-channel configuration, the selected frequencies are
\[
\begin{aligned}
f_k=\{&
9.684,\;9.716,\;9.748,\;9.780,\\
&
9.810,\;9.842,\;9.874,\;9.906,\\
&
9.938,\;9.970,\;10.002,\;10.034,\\
&
10.064,\;10.096,\;10.128,\;10.160
\}\ {\rm GHz}.
\end{aligned}
\]
The five-channel configuration used for the reported XOR result was
selected retrospectively from the tested channel counts by
maximizing the smaller of the two grouped-validation accuracies
($x$-group and mirror-$y$), with ties resolved in favor of the
smaller $N_b$. The reported accuracy should therefore be interpreted
as a characterization of the selected configuration rather than as
an independent post-selection test estimate.

\subsection{Shared-weight two-branch readout}

The proposed experimental model combines the two measured branches
using a shared set of spectral weights,
\begin{equation}
    z(x,y)
    =
    b
    +
    \sum_{k=1}^{N_b}
    w_k
    \left[
        P^{(-)}(x,y;f_k)
        -
        P^{(+)}(x,y;f_k)
    \right].
    \label{eq:sm_shared_readout}
\end{equation}

This shared-weight construction uses the same physical spectral
mapping for the two mirror-addressed branches and differs from a
general two-branch linear readout with independently adjustable
weights.

For comparison, we additionally evaluate an independent-weight
two-branch model,
\begin{equation}
    z_{\mathrm{ind}}
    =
    b+
    \sum_k w_k^{(+)}P^{(+)}(f_k)
    +
    \sum_k w_k^{(-)}P^{(-)}(f_k),
\end{equation}
as well as a single-branch resonant model
\begin{equation}
    z_{\mathrm{single}}
    =
    b+
    \sum_k w_kP^{(+)}(f_k),
\end{equation}
and an affine input baseline
\begin{equation}
    z_{\mathrm{lin}}
    =
    b+w_xx+w_yy .
\end{equation}

\subsection{Training procedure}

All model features are standardized using the mean and standard
deviation calculated from the corresponding training set only.

The weights appearing in Eqs.~(69)--(71) are reported in the
original power coordinates. If a feature is standardized as
$\widetilde X_k=(X_k-\mu_k)/\sigma_k$, the fitted coefficients are
converted back according to
\begin{equation}
w_k=\frac{\widetilde w_k}{\sigma_k},
\qquad
b=\widetilde b-\sum_k w_k\mu_k .
\end{equation}

For XOR classification, the linear readout weights are trained using
$L_2$-regularized logistic regression. The inverse regularization
strength is selected from
\begin{equation}
    C\in
    \{
    10^{-3},10^{-2},10^{-1},1,10,10^2,10^3
    \}.
\end{equation}
The optimization uses the \texttt{lbfgs} solver with a maximum of
2500 iterations and a tolerance of $10^{-9}$. The regularization
parameter is selected by grouped five-fold validation within the
training set according to classification error. In the case of a
tie, the smaller value of $C$ is selected. The final decision rule
assigns class 1 when the logistic decision score satisfies $z>0$.

For continuous regression of $g=-xy$, the readout is trained using
ridge regression with
\begin{equation}
    \alpha
    \in
    \{
    10^{-4},10^{-3},10^{-2},10^{-1},
    1,10,10^2,10^3
    \}.
\end{equation}
The regularization strength is selected by grouped five-fold
validation according to mean-squared error; ties are resolved in
favor of the larger value of $\alpha$.

The classification and continuous-regression weights are trained
independently. Consequently, the regression $R^2$ values should not
be interpreted as a goodness-of-fit measure of the classification
decision score.

\subsection{Quadratic form of the shared resonant readout}

To examine the physical form of the outer mapping learned by the
two-branch model, we freeze the shared spectral weights obtained
from the continuous regression of $g=-xy$ and inspect the response
of a single branch. For the 16-channel model, the frozen
single-branch output is
\begin{equation}
h(x,y)
=
\sum_{k=1}^{16}
w_k P(x,y;f_k),
\label{eq:sm_single_branch_output}
\end{equation}
where the weights $w_k$ are those of the shared-weight model and
are not retrained for the following analysis. A branch-independent
bias is omitted because it cancels when the two branches are
subtracted.

We then compare $h$ with the independently extracted resonance
frequency $f_r(x,y)$. For numerical conditioning, the frequency is
written as
\begin{equation}
u
=
\frac{f_r-\bar f_r}{0.1~\mathrm{GHz}},
\end{equation}
For the full-grid analysis, $\bar f_r$ denotes the mean resonance
frequency over all measured positions. For grouped evaluation,
$\bar f_r$ is computed using the training positions of each fold
only.
and the frozen branch response is fitted by
\begin{equation}
h(u)
=
c_0+c_1u+c_2u^2 .
\label{eq:sm_quadratic_readout}
\end{equation}

Over the complete $20\times20$ measured grid, the quadratic model
gives
\begin{equation}
R^2=0.9401,
\end{equation}
whereas a linear fit gives only $R^2=0.0142$. The quadratic
dependence is also preserved under grouped evaluation: fitting the
polynomial using only the training positions gives test
$R^2=0.9361$ for the $x$-group split and $R^2=0.9383$ for the
mirror-$y$ split.

As an additional consistency check, replacing the measured
single-branch readout by the fitted quadratic function and then
subtracting the two mirror-addressed branches reproduces most of
the actual shared-weight difference output, with $R^2=0.9597$ on
the complete grid. These results show that the shared resonant
readout learned for the continuous product task is well
approximated by a quadratic function of the resonance frequency
over the experimentally sampled range.

For grouped evaluation, the reported coefficient of determination
is computed as
\begin{equation}
R^2
=
1-
\frac{\sum_s \mathrm{SSE}_s}
{\sum_s \mathrm{SST}_s},
\end{equation}
where $s$ indexes the outer folds and each $\mathrm{SST}_s$ is
evaluated within the corresponding test fold.

\subsection{Mirror-pair grouped cross-validation}

The mirror-addressed construction in
Eq.~(\ref{eq:sm_mirror_branch}) introduces a potential data-leakage
path if a spectrum at $(x,y)$ is included in the training set while
its mirrored counterpart $(x,-y)$ appears in the test set. We
therefore use grouped cross-validation in which mirrored $y$
positions are always assigned to the same fold.

For the sorted position index $j=0,\ldots,19$, the group label is
defined as
\begin{equation}
    G_j=\min(j,19-j),
\end{equation}
resulting in ten mirror-pair groups.
A five-fold grouped split is then performed. Each outer fold leaves
out two complete mirror-pair groups, corresponding to four $y$
positions and all twenty $x$ positions. Each fold therefore
contains 320 training samples and 80 test samples.

All frequency selection, feature normalization, and regularization
selection are performed using the training portion of the
corresponding fold. The final reported predictions are the
out-of-fold predictions collected from all five folds, so that each
of the 400 physical input combinations appears exactly once in the
test set.

The same mirror-pair grouping is used for the inner validation
performed during hyperparameter selection.

As an additional robustness test, we also use an $x$-group split,
in which each outer fold holds out four complete $x$ positions and
all corresponding $y$ positions. Each fold again contains 320
training samples and 80 test samples.

\subsection{XOR classification and continuous regression}

Using five fixed-frequency channels, the shared-weight two-branch
model correctly classifies 394 of the 400 out-of-fold samples,
corresponding to an XOR classification accuracy of
\begin{equation}
    98.5\%.
\end{equation}

The corresponding confusion matrix is
\begin{equation}
    \begin{pmatrix}
        197 & 3\\
        3 & 197
    \end{pmatrix},
\end{equation}
where rows denote the true classes and columns the predicted
classes.

Under the same grouped evaluation, the affine input baseline,
single-branch resonant model, shared-weight two-branch model, and
independent-weight two-branch model achieve five-channel
classification accuracies of
$50.0\%$, $83.75\%$, $98.5\%$, and $99.0\%$, respectively.

The close performance of the shared- and independent-weight
two-branch models indicates that allowing the two branches to use
independently optimized spectral weights provides only a small
classification advantage for the present data.

We additionally evaluate the continuous target $g=-xy$.
For five readout channels, the shared-weight two-branch model gives
an out-of-fold coefficient of determination
\begin{equation}
    R^2=0.7751.
\end{equation}
Increasing the readout bank to sixteen frequency channels improves
the continuous regression to
\begin{equation}
    R^2=0.98047,
\end{equation}
with an NRMSE of $0.02574$. Here the NRMSE is defined as the
root-mean-square error divided by the width of the predefined target
range $[-1,1]$.

For comparison, the sixteen-channel single-branch model reaches
$R^2=0.69293$, whereas the independent-weight two-branch model gives
$R^2=0.98052$. The difference between the shared- and
independent-weight two-branch models is therefore approximately
$5\times10^{-5}$ in $R^2$.

The high classification accuracy obtained with five channels and
the lower five-channel continuous-regression $R^2$ are not
contradictory. XOR classification requires primarily the correct
sign of the nonlinear output, whereas continuous regression
additionally requires accurate reconstruction of its magnitude.

\subsection{Scope of the experimental demonstration}

The present experiment should be interpreted as a proof-of-principle
construction of a two-branch resonant KAN from measured
single-cavity responses and mirror-input addressing. Only one
physical cavity containing two movable perturbations is measured;
the second branch is generated from the measured mirror input
according to Eq.~(\ref{eq:sm_mirror_branch}).

Furthermore, the classification and regression models operate
directly on the experimentally measured fixed-frequency power
responses. They therefore do not assume an ideal Lorentzian
lineshape, a constant quality factor, or an exactly additive
resonance shift.

The observed XOR and continuous-product performance demonstrates
that the measured resonant basis supports the required two-branch
nonlinear mapping. It does not, by itself, imply that each
experimental branch realizes an exact quadratic function.

\bibliography{references.bib}

%% file: chapters/Introduction.tex
\textit{Introduction.---}
Optical neural networks offer a promising route toward high-throughput and energy-efficient
information processing \cite{shastri_photonics_2021,wetzstein_inference_2020,hamerly_large-scale_2019,feldmann_parallel_2021}, exploiting interference or diffraction
to perform large-scale linear transformations \cite{shen_deep_2017,lin_all-optical_2018,zhou_large-scale_2021,xu_taichi_2024}.
Their nonlinear operations commonly rely on material optical nonlinearities \cite{zuo_all-optical_2019,miscuglio_all-optical_2018,wu_field-programmable_2025}, 
electro-optic modulation \cite{williamson_reprogrammable_2020,huang_programmable_2022,bandyopadhyay_single-chip_2024}, or optical--electrical--optical 
conversion \cite{wang_chip-based_2022,tait_silicon_2019,ashtiani_on-chip_2022}.
However, these approaches require additional power, control, and hardware \cite{wanjura_fully_2024,xia_nonlinear_2024}.
An alternative route is to encode the input variables into the physical parameters
of a linear wave system.
The wave dynamics remain linear with respect to the field amplitude, while the scattering 
response depends nonlinearly on the encoded input.
This mechanism, termed structural nonlinearity \cite{eliezer_tunable_2023}, enables nonlinear computation without 
requiring intrinsic physical nonlinearities.

Structural nonlinearity can be enhanced when the propagating wave repeatedly encounters 
the input-encoded structure.
These repeated interactions, connected by linear propagation, generate higher-order terms 
in the input variables.
This principle has been exploited to realize nonlinear optical transformations using linear 
optical systems \cite{yildirim_nonlinear_2024}.
However, for finite-depth repeated-encoding architectures, the accessible polynomial order 
is limited by the number of input repetitions \cite{yildirim_nonlinear_2024,wanjura_fully_2024}.
The realization of higher polynomial orders therefore requires additional input repetitions.
Recurrent scattering provides an intrinsic mechanism for accumulating repeated interactions 
without additional encoding stages.

Cavities inherently support recurrent scattering \cite{xia_nonlinear_2024}, allowing 
repeated interactions with the same input-encoded structure.
In a resonant cavity, these repeated scattering events interfere coherently and shape 
the modal response.
Consequently, a cavity that remains linear with respect to the field amplitude can exhibit 
a strongly nonlinear scattering response to the input encoded in the cavity structure.
Previous studies have exploited multiple-scattering cavities for structural
nonlinearity and nonlinear information processing
\cite{eliezer_tunable_2023,xia_nonlinear_2024},
while modal descriptions of linear resonant systems have further shown how detunings and
couplings can generate nonlinear computational mappings
\cite{wanjura_fully_2024}.
Unlike finite-depth repeated encoding, resonant recurrence resums
these repeated interactions into a non-polynomial modal response
without requiring additional encoding stages.

This resonant response also provides a natural compositional
structure for computation.
An input encoded into the cavity parameters modifies the resonant modes,
while the corresponding spectral response provides a nonlinear readout of these changes.
This two-stage mapping matches the functional composition of a
Kolmogorov--Arnold network (KAN)
\cite{kolmogorov_representation_1957,arnold_functions_1957,liu_kan_2025}.
Recent work has implemented KAN-like nonlinear mappings in linear optical systems
through repeated data encoding, with higher polynomial orders obtained by increasing
the encoding depth \cite{stroev_programmable_2026}.
Resonant systems offer a complementary route:
their modal response can provide non-polynomial mappings without increasing the number 
of encoding stages.
This raises the central question of how the underlying resonance physics
determines the realizable function family and its representation capacity.

Here, we show that recurrent scattering in resonant cavities
provides a route to non-polynomial computation, with the
Kolmogorov--Arnold representation used to organize the resulting
physical mappings.
Structural perturbations and resonant spectral responses naturally
realize the inner and outer functions of the K--A representation.
Numerically, we identify the resonance linewidth and branch number
as complementary controls of representation capacity, while a
microwave experiment validates the two-stage mapping through XOR
classification.
Together, these results connect recurrent resonance physics to a
controllable family of nonlinear computational mappings in linear
wave systems.

%% file: chapters/framework_2.tex
\begin{figure*}[t]
    \centering
    \includegraphics[width=0.95\textwidth]{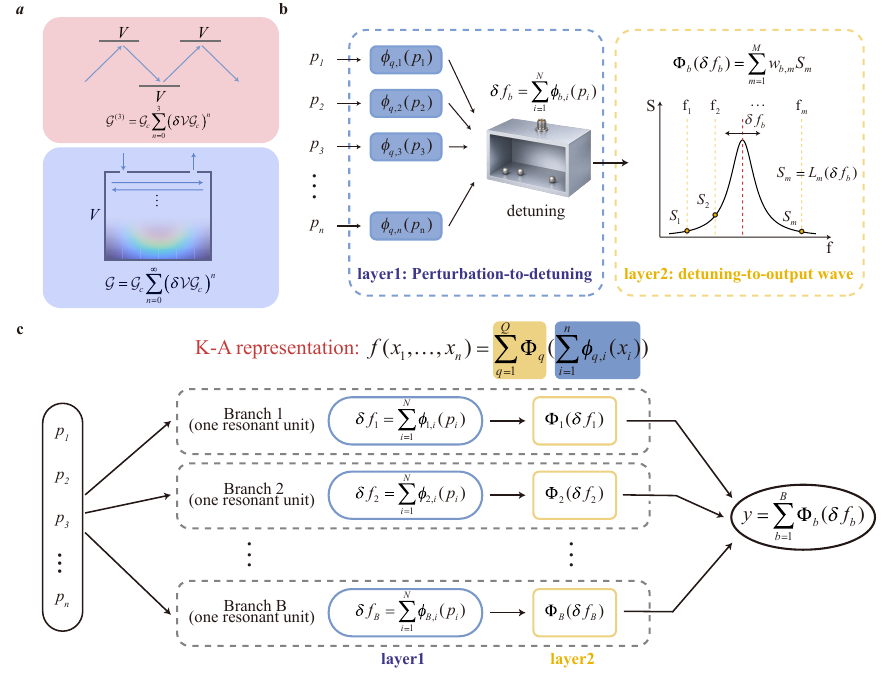}
    \caption{
    (a) Comparison between finite multiple scattering and recurrent scattering in a resonant cavity.
    (b) Theoretical mapping of a single resonant branch (resonance frequency $f_r=\omega_r/(2\pi)$).
    (c) Multiple resonant branches are combined to realize the Kolmogorov--Arnold form with each cavity implementing one compositional branch.
    }
    \label{fig:framework}
\end{figure*}

\textit{Design principles.---}
The resonant cavity provides recurrent propagation that repeatedly
returns the wave to the same input-encoded structure.
As shown in Fig.~\ref{fig:framework}(a), let $\mathcal{G}_{c}(\omega)$ denote the Green operator of the
unperturbed cavity and
$\delta\mathcal{V}(\mathbf{x})$ the structural perturbation
controlled by the input variables.
The perturbed Green operator satisfies the Dyson equation
\begin{equation}
    \mathcal{G}
    =
    \mathcal{G}_{c}
    +
    \mathcal{G}_{c}
    \delta\mathcal{V}
    \mathcal{G}.
    \label{eq:dyson}
\end{equation}
Formally iterating Eq.~(\ref{eq:dyson}) gives
\begin{equation}
    \mathcal{G}
    =
    \mathcal{G}_{c}
    +
    \mathcal{G}_{c}\delta\mathcal{V}\mathcal{G}_{c}
    +
    \mathcal{G}_{c}\delta\mathcal{V}
    \mathcal{G}_{c}\delta\mathcal{V}\mathcal{G}_{c}
    +\cdots ,
    \label{eq:recurrent_series}
\end{equation}
Each higher-order term represents an additional encounter with 
the same input-encoded perturbation.

Near an isolated cavity resonance, the response is dominated by a 
single quasinormal mode. 
The unperturbed Green operator can therefore be approximated as 
\begin{equation} 
    \mathcal{G}_{c}(\omega) 
    \simeq 
    \frac{\mathcal{R}_{0}} {\omega-\widetilde{\omega}_{0}}, 
    \qquad 
    \widetilde{\omega}_{0} = \omega_{0}-i\gamma_{0}, 
    \label{eq:unperturbed_pole} 
\end{equation} 
where $\mathcal{R}_{0}$ denotes the contribution of the dominant mode. 
The input-dependent structural perturbation $\delta\mathcal{V}(\mathbf{x})$ 
shifts this complex resonance frequency by $\delta\widetilde{\omega}(\mathbf{x})$. 
For weak perturbations, the resulting resonance shift can be described by first-order 
cavity perturbation theory, as introduced below.

With this single-mode description, each insertion of $\delta\mathcal{V}$ in 
the repeated-interaction series contributes the same effective modal 
perturbation $\delta\widetilde{\omega}$. 
The resonant part of the series can therefore be written as 
\begin{equation} 
    \mathcal{G}
    \propto \frac{1}{D} + 
    \frac{\delta\widetilde{\omega}}{D^{2}} + 
    \frac{\delta\widetilde{\omega}^{\,2}}{D^{3}} +\cdots , 
    \qquad D=\omega-\widetilde{\omega}_{0}. 
    \label{eq:projected_series} 
\end{equation}
which resums to
\begin{equation}
    \mathcal{G}
    \propto
    \frac{1}
    {
        \omega
        -
        \widetilde{\omega}_{0}
        -
        \delta\widetilde{\omega}(\mathbf{x})
    }.
    \label{eq:pole_resummation}
\end{equation}
Thus, the input-encoded structural perturbation is converted by the
recurrent resonant dynamics into an input-dependent displacement of
the cavity pole.

For excitation through port 1 and readout at port 2,
the transmission coefficient inherits this pole,
\begin{equation}
    S_{21}(\omega,\mathbf{x})
    \simeq
    S_{\mathrm{bg}}(\omega,\mathbf{x})
    +
    \frac{A(\mathbf{x})}
    {
        \omega-\widetilde{\omega}_{0}
        -\delta\widetilde{\omega}(\mathbf{x})
    },
    \label{eq:single_pole}
\end{equation}
where $A$ describes the modal excitation and readout coupling,
and $S_{\mathrm{bg}}$ contains the nonresonant contribution.

We next relate the pole shift to the encoded variables.
In the weak-perturbation regime considered here, we retain
the real resonance-frequency shift
$\Delta\omega=\mathrm{Re}\,\delta\widetilde{\omega}$.
For a weak and localized perturbation, first-order cavity
perturbation theory gives
\begin{equation}
    \Delta\omega_i
    \simeq
    C_i
    \int_{\Omega_i}
    g_i(\mathbf{r})\,dV ,
    \label{eq:perturbation_shift}
\end{equation}
where $\Omega_i$ denotes the volume of the $i$th perturbing element 
(see Supplemental Material for the explicit form used here).
As the perturbation moves along the trajectory $\mathbf{r}_i(x_i)$, 
where $x_i$ specifies its position along the trajectory, the resulting resonance shift defines a univariate mapping,
\begin{equation}
    \phi_i(x_i)
    \equiv
    \Delta\omega_i
    \left[
        \mathbf{r}_i(x_i)
    \right].
    \label{eq:inner_function}
\end{equation}
For sufficiently weak perturbations, the first-order resonance shifts
are approximately additive,
\begin{equation}
    u(\mathbf{x})
    \equiv
    \Delta\omega(\mathbf{x})
    \simeq
    \sum_i\phi_i(x_i).
    \label{eq:additive_shift}
\end{equation}

Neglecting the variation of the resonance linewidth and the nonresonant background 
(see Supplemental Material for details), the transmission intensity at a fixed 
readout frequency $\omega_m$ takes the Lorentzian form 
\begin{equation} 
    L_m(u) 
    \propto 
    \frac{1} 
    { [\omega_m-\omega_0-u]^2 + \gamma_0^2 }, 
    \label{eq:lorentzian_basis} 
\end{equation} 
where $u(\mathbf{x})\simeq\sum_i\phi_i(x_i)$ is the input-induced resonance shift.

Equation~(\ref{eq:lorentzian_basis}) therefore provides a
non-polynomial mapping of the input-induced resonance shift,
whose local expansion contains terms of arbitrarily high order
in $u$ (see Supplemental Material).

Sampling the shifted resonance at a set of fixed frequencies
$\{\omega_m\}_{m=1}^{M}$ provides a family of nonlinear spectral
basis functions.
Their weighted combination defines an outer univariate mapping,
\begin{equation}
    \Phi(u)
    =
    \sum_{m=1}^{M}
    w_m L_m(u).
    \label{eq:outer_function}
\end{equation}
Combining this spectral mapping with the spatially encoded
frequency shifts gives
\begin{equation}
    F(\mathbf{x})
    =
    \Phi
    \left[
        \sum_i\phi_i(x_i)
    \right].
    \label{eq:resonance_primitive}
\end{equation}
As shown in Fig.~\ref{fig:framework}(b), the modal perturbation landscape therefore implements the inner
univariate mappings $\phi_i$, whereas the resonant spectral response
implements the outer mapping $\Phi$.
Equation~(\ref{eq:resonance_primitive}) has the compositional form of
one branch of the Kolmogorov--Arnold representation
\cite{kolmogorov_representation_1957,
arnold_functions_1957,
liu_kan_2025}.
Multiple resonant modules provide multiple such branches and can be
combined to form the corresponding KAN architecture, as shown in Fig.~\ref{fig:framework}(c).

\begin{figure}[t]
    \centering
    \includegraphics[width=0.48\textwidth]{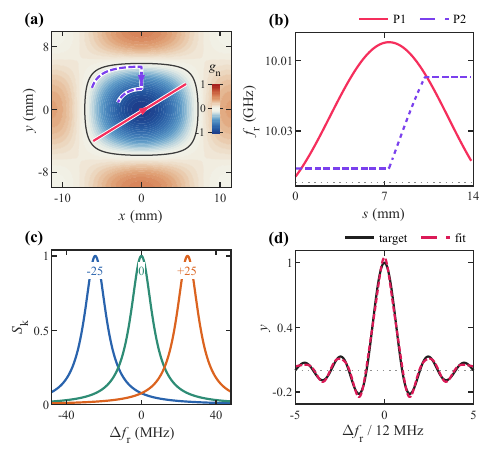}
    \caption{
    Spatial and spectral construction of a resonant KAN primitive.
    (a) Normalized spatial perturbation landscape
    $g_n=g/\max|g|$ in a rectangular cavity with a resonance frequency of approximately
    $10~\mathrm{GHz}$, together with two representative perturbation trajectories, P1 and P2.
    (b) Path-dependent resonance frequency $f_r$ as a function of displacement $s$ along P1 and P2.
    (c) Representative resonant spectral basis functions sampled at fixed
    readout frequencies. Fifteen basis functions are used for the fitting, with three representative
    ones shown for clarity.
    (d) Synthesis of a target sinc function by a weighted combination of the spectral basis functions.
}
    \label{fig:primitive}
\end{figure}

The physical ingredients underlying
Eq.~(\ref{eq:resonance_primitive})
are illustrated in Fig.~\ref{fig:primitive}.
Figure~\ref{fig:primitive}(a) shows the normalized spatial perturbation
landscape
$g_n=g/\max|g|$
together with two representative perturbation trajectories,
P1 and P2.
Because the same modal landscape is sampled along different paths,
the two trajectories generate distinct one-dimensional frequency mappings.
This is shown in Fig.~\ref{fig:primitive}(b),
where the resonance frequency varies differently with the displacement
along P1 and P2.
The result demonstrates that the inner function $\phi_i$
can be physically shaped through the perturbation trajectory
without altering the underlying resonant structure.

The second stage is provided by the resonant spectral response.
As shown in Fig.~\ref{fig:primitive}(c),
a perturbation shifts the resonance relative to several fixed
readout frequencies.
Each readout channel therefore samples a different Lorentzian function
of the resonance shift,
providing a set of nonlinear spectral basis functions.
By optimizing their linear weights,
these responses can be combined to synthesize a desired outer function.
Figure~\ref{fig:primitive}(d) illustrates this principle by fitting
a target sinc function with a weighted combination of the calculated
spectral responses.

These results establish the two physical degrees of freedom underlying
the resonant functional primitive:
the perturbation trajectory determines the inner input--frequency mapping,
while the resonant spectrum determines the available outer-function basis.
The next question is therefore how the resonance linewidth sets the functional 
richness within each branch, and how combining multiple branches extends the 
representation capacity at the network level.

%% file: chapters/capacity.tex
\begin{figure}[t]
    \centering
    \includegraphics[width=\columnwidth]{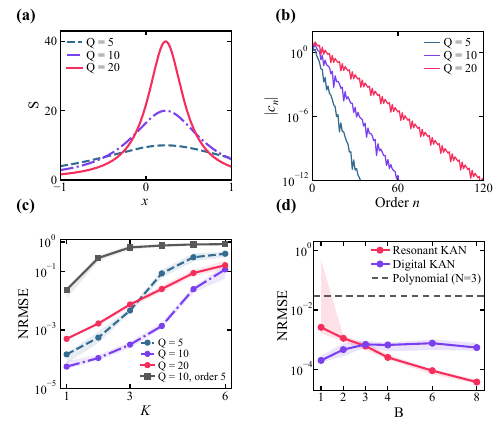}
    \caption{\textbf{Representation capacity of resonant KANs.}
    (a) Resonant response for different quality factors $Q$.
    (b) Chebyshev coefficients of these responses.
    Larger $Q$ produces a more slowly decaying high-order envelope.
    (c) Test NRMSE for random Fourier targets of complexity $K$
    using 15 fixed spectral channels.
    The $T_5$ reference is obtained by truncating the Chebyshev
    expansion of each $Q=10$ resonant basis function at fifth
    order, preserving the same 15 readout channels while removing
    the higher-order components.
    (d) Test NRMSE versus branch number $B$ for
    $f(x,y)=(x+y)/(1+xy)$,
    comparing resonant and digital spline KANs with
    the same $[2,B,1]$ topology.
    The dashed line denotes a total-degree-three polynomial fit.
    Lines and shaded bands indicate the median and interquartile
    range over 20 targets in (c) and 10 initializations in (d).
    All results are numerical.}
    \label{fig:capacity}
\end{figure}

\textit{Representation capacity.---}
The compositional structure established above suggests two
levels of representation capacity: the functional richness
within each resonant branch and the broader function space
obtained by combining multiple branches.
The former is governed by the resonant spectral response,
whereas the latter grows through the addition of independent
compositional branches.
We therefore examine these two levels separately:
the single-branch capacity in
Fig.~\ref{fig:capacity}(a--c)
and the network-level capacity in
Fig.~\ref{fig:capacity}(d).
Numerical details are provided in the Supplemental Material.

To isolate the spectral contribution, we consider a trajectory
that produces a linear resonance shift,
$\Delta \omega[\mathbf r(x)] = \alpha x $.
The resulting functional complexity therefore originates
from the resonant response itself.
Increasing $Q$ reduces the resonance linewidth and enhances
the spectral variation of the response \cite{fan_temporal_2003}
[Fig.~\ref{fig:capacity}(a)].
To quantify how the resonant response is distributed across
different polynomial orders, we expand it in a Chebyshev basis,
\begin{equation}
    S(x)=\sum_{n=0}^{\infty} c_n T_n(x).
\end{equation}
The high-order coefficients decay more slowly as $Q$ increases
[Fig.~\ref{fig:capacity}(b)], indicating a broader distribution
over polynomial orders.

We therefore examine how the higher-order content of the resonant
response affects representation capacity using random Fourier targets,
\begin{equation}
    f_K(x)=\sum_{k=1}^{K}
    \left[
        a_k\cos(k\pi x)+b_k\sin(k\pi x)
    \right],
\end{equation}
with $K$ controlling the target complexity.
For all $Q$, the number and positions of the 15 readout channels
are kept fixed.
As a finite-order reference, we additionally truncate the
Chebyshev expansion of each of the 15 $Q=10$ resonant basis
functions at $T_5$, thereby removing all Chebyshev components above 
fifth order while preserving the same spectral channels.

The full $Q=10$ basis outperforms its $T_5$-truncated
counterpart across the tested range
[Fig.~\ref{fig:capacity}(c)], showing that components above
fifth order contribute to the representation of complex targets.
For the full resonant bases, the best performance occurs at an
intermediate $Q$: increasing $Q$ enriches the higher-order content
but simultaneously narrows the resonances sampled by the fixed
readout frequencies.
Representation therefore depends on a balance between functional
richness and spectral coverage.

We next examine the network-level representation capacity
obtained by combining multiple resonant branches.
We use a dimensionless regression target drawn from the Feynman
Symbolic Regression Database \cite{udrescu_ai_feynman_2020}.
The target function is
\begin{equation}
    f(x,y)=\frac{x+y}{1+xy}
    =\tanh\!\left[
        \operatorname{atanh}(x)+\operatorname{atanh}(y)
    \right].
\end{equation}
This identity directly matches the structure of a single
K--A branch: the two inputs are independently transformed by
$\operatorname{atanh}(\cdot)$, summed, and then passed through
$\tanh(\cdot)$.
The digital KAN can approximate the required
$\operatorname{atanh}$ and $\tanh$ mappings with flexible spline
edges \cite{liu_kan_2025}, and therefore already reaches a low
error with a single branch.
A resonant branch is more constrained: its inner mappings are
set by perturbation trajectories and its outer mapping by a finite
spectral basis.
The restricted function family of a single physical branch
can be expanded by combining multiple cavity branches.
The resulting network is
\begin{equation}
    \hat f(x,y)=b_0+\sum_{q=1}^{B}
    \Phi_q\!\left[
        \phi_{q,x}(x)+\phi_{q,y}(y)
    \right],
\end{equation}
where each additional cavity contributes an independently
trainable branch.
As $B$ increases, the median NRMSE decreases from approximately
$2.6\times10^{-3}$ at $B=1$ to $3.8\times10^{-5}$ at $B=8$
[Fig.~\ref{fig:capacity}(d)].

These results identify two complementary controls of representation
capacity: $Q$ governs the functional richness within each resonant
branch, while increasing $B$ expands the accessible function space
by combining multiple resonant primitives.

%% file: chapters/experiment.tex
\begin{figure}[t]
    \centering
    \includegraphics[width=\columnwidth]{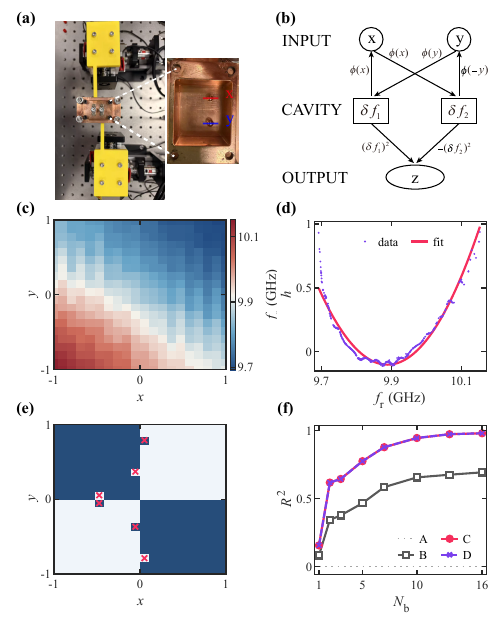}
    \caption{
        Experimental validation of a $[2,2,1]$ KAN using single-cavity
        measurements and mirror-input addressing.
        (a) Experimental setup and cavity interior.
        (b) Mirror-addressed two-branch architecture with a shared
        spectral readout.
        (c) Resonance frequency over the $20\times20$ input grid.
        (d) Frozen 16-channel single-branch readout and quadratic fit
        ($R^2=0.940$, full grid).
        (e) Five-channel XOR predictions under mirror-pair grouped
        cross-validation: dark/light regions indicate classes 1/0,
        and red crosses mark errors (accuracy $98.5\%$).
        (f) Out-of-fold $R^2$ for the continuous target $g=-xy$
        versus the number of readout channels $N_b$ for the linear
        baseline (A), single-branch resonant model (B), shared-weight
        two-branch model (C), and independent-weight two-branch model (D).
    }
    \label{fig:experimental_validation}
\end{figure}

\textit{Experimental validation.---}
To experimentally validate the theoretical model, we construct
a two-branch KAN mapping from measured single-cavity responses
using mirror-input addressing.
We consider the XOR task for continuous inputs
$x,y\in[-1,1]$, with $t=1$ for $xy<0$ and $t=0$ for $xy>0$.
Since the two classes occupy alternating quadrants, they are not
linearly separable.
A $[2,2,1]$ KAN can realize this mapping using two branches
associated with $x+y$ and $x-y$, followed by a quadratic
outer mapping:
\begin{equation}
    (x+y)^2-(x-y)^2=4xy.
    \label{eq:xor_square_identity}
\end{equation}
The sign of the output therefore determines the XOR class.
We use the equivalent target $g=-xy$, for which positive
values are assigned to class 1.

Figure~\ref{fig:experimental_validation}(a) shows the experimental
platform, consisting of a microwave resonant cavity with two movable
perturbing elements. 
The red and blue trajectories encode the inputs $x$ and $y$,
respectively.
We measure $S_{21}$ over a complete $20\times20$
joint-position grid and normalize the two input coordinates
to $[-1,1]$.
Although only one physical cavity is measured, the second
KAN branch is emulated by reversing the trajectory assigned
to the input $y$.
The two branches are therefore represented by
$P(x,y;f)$ and $P(x,-y;f)$, respectively
[Fig.~\ref{fig:experimental_validation}(b)].
Their shared-weight output is
\begin{equation}
z(x,y)=b+\sum_{k=1}^{N_b}w_k
\left[
P(x,-y;f_k)-P(x,y;f_k)
\right],
\label{eq:experimental_readout}
\end{equation}
where $P=|S_{21}|^2$ and both branches use the same
readout frequencies and weights.
Thus, the two-branch computation is constructed from measured
cavity responses, with the second branch realized by reversing
the $y$-encoding trajectory rather than by using a second
physical cavity.

Figures~\ref{fig:experimental_validation}(c) and
\ref{fig:experimental_validation}(d) verify the two mappings
required by the $[2,2,1]$ architecture.
The joint resonance-frequency map provides the inner
sum- and difference-like coordinates through the original
and reversed $y$ encodings
[Fig.~\ref{fig:experimental_validation}(c)].
A weighted combination of 16 frequency channels produces an
approximately quadratic dependence on the resonance shift
[Fig.~\ref{fig:experimental_validation}(d)], providing the
quadratic outer mapping required by the XOR construction.
The quadratic fit gives $R^2=0.940$ over the measured grid.
The correspondence is not exact because the measured frequency
shift is neither perfectly linear along the perturbation
trajectories nor exactly additive between the two perturbations
(see Supplemental Material).

Figure~\ref{fig:experimental_validation}(e) shows the XOR
classification obtained with five frequency channels.
Using nest five-fold cross-validation with mirrored $y$ positions
kept in the same fold, the measured responses achieve an
accuracy of $98.5\%$, with only six errors among the 400
input combinations.

We further test the underlying continuous mapping by fitting
the target $g=-xy$.
As the number of frequency channels increases from 5 to 16,
the test $R^2$ rises from $0.775$ to $0.980$
[Fig.~\ref{fig:experimental_validation}(f)].
The shared-weight result is nearly identical to that obtained
with independently optimized branch weights, whereas the
single-branch model reaches only $R^2=0.693$.
These results experimentally verify both the XOR classification
and the underlying nonlinear mapping $g=-xy$ of the proposed
two-branch construction.

Details of the experimental setup, data processing,
frequency-channel selection, and cross-validation procedures
are provided in the Supplemental Material.